**Eating Smart: Free-ranging dogs follow an optimal foraging strategy while scavenging in groups**

Rohan Sarkar[1], Sreelekshmi R[1,2], Abhijit Nayek[1,3,4], Anirban Bhowmick[1,5], Poushali Chakraborty[1,6,7], Rituparna Sonowal[1,8], Debsruti Dasgupta[1,9,10], Rounak Banerjee[1,9], Aritra Roy[1,11,12], Amartya Baran Mandal[1,13,14], and Anindita Bhadra[1*]

1. Department of Biological Sciences, Indian Institute of Science Education and Research-Kolkata, West Bengal, India, 741246
2. Centre for Systems and Computational Biology, Hyderabad Central University, Hyderabad, Telangana, India, 500046
3. Department of Biomedical Science, Acharya Narendra Dev College, University of Delhi, New Delhi, Delhi, India, 110019
4. School of Biosciences and Bioengineering, Indian Institute of Technology- Mandi, Himachal Pradesh, India, 175005 (Present Address)
5. Department of Biological Sciences, Indian Institute of Science Education and Research, Berhampur, Odisha, India, 760010
6. Department of Life Science and Biotechnology, Jadavpur University, Kolkata, West Bengal, India, 700032
7. Department of Chemical Engineering, Jadavpur University, Kolkata, West Bengal, India, 700032 (Present Address)
8. Department of Animal and Food Sciences, Texas Tech University, Lubbock, Texas, USA, 79409 (Present Address)
9. Department of Biotechnology, St. Xavier's College (Autonomous), Kolkata, West Bengal, India, 700016
10. MBA- Business Management, Xavier Institute of Management, Bhubaneswar, Odisha, India, 751013 (Present Address)
11. Department of Energy and Environment, TERI School of Advanced Studies, New Delhi, Delhi, India, 110070
12. Project Tiger- AITE, Wildlife Institute of India, Dehradun, Uttarakhand, India, 248001, (Present Address)
13. Department of Zoology, Ramakrishna Mission Vivekananda Centenary College, Rahara, Kolkata, West Bengal, India, 700118
14. Department of Biochemical Engineering, University of Debrecen, Debrecen, Hungary, H-4032

* Corresponding Author

Dr. Anindita Bhadra

Behaviour and Ecology Lab, Department of Biological Sciences, Indian Institute of Science Education and Research Kolkata, Mohanpur Campus, Mohanpur, Nadia, West Bengal, India, Pin: 741246

Phone no: +91 33 6136 0000 ext 1223

E-mail: abhadra@iiserkol.ac.in

**Abstract**

Foraging and acquiring of food is a delicate balance between managing the costs (both energy and social) and individual preferences. Previous research on the solitary foraging of free-ranging dogs showed that they prioritized the nutritionally highest valued food patch first but don't ignore other less valuable food either, displaying typical scavenger behaviour. The current experiment was carried out on groups of dogs with the same set-up to see the change in foraging strategies, if any, under the influence of social cost like intra-group competition. We found multiple differences between the strategies of dogs foraging alone versus in groups with competition playing an implicit role in the dogs' decision making when foraging in groups. Dogs were able to continually assess and evaluate the available resources in a patch and adjust their behaviour accordingly. Foraging in groups also provided benefits of reduced individual vigilance. The various decisions and choices made seemed to have a basis in the optimal foraging theory wherein the dogs harvested the nutritionally richest patch possible with the least risk and cost involved but was willing to compromise if that was not possible. This underscores the cognitive, quick decision-making abilities and adaptable behaviour of these dogs.

**Significance Statement**

The purpose of this study is to explore an aspect of ecoethology of free-ranging dogs, that of their scavenging behaviour. Despite their ubiquitous presence on Indian streets, not much is known about these dogs. This study is the first of its kind to explore in detail the foraging strategies of these dogs under different social conditions. We showed experimentally by simulating real world conditions that these dogs are quick thinkers and behaviourally flexible in their strategies displaying qualities of an efficient scavengers. We also found that these strategies show evidence of an optimal foraging model.

**Introduction**

Animal foraging strategies are a delicate balance between food preference and food selection influenced by various factors like nutritional requirements, availability of food, predation, competition etc[1]. Food preference is defined as the discrimination exerted by animals on different food types resulting in the selection of one food type over the other when no constraints bear on their choice[2]. Multiple experiments have shown that, given the chance, animals do indeed prefer certain food items more than others[3,4]. But such idealized situations rarely occur in the natural world and thus animals may not always be able to acquire food according to their preference. For example, goats and giant anteaters make different dietary choices when foraging in wild versus in experimental conditions[5,6]. Thus, it is necessary to distinguish between food preference and food selection, defined as preference modified by environmental conditions[7]. The search for the best possible food with least risk and energy output gives rise to different strategies. Members of a group-living species might forage alone or in groups depending on the food abundance in the area[8]. Foraging in groups provide animals with the advantages of lower predation risk and higher efficiency[9]. But living in groups have their disadvantages too. Intra-group competition is one of the costs of group living[10]. Competition has an effect on the quality and quantity of food acquired by individuals. As a trade-off to competition, individuals may change dietary preferences or forage from lower quality patches[11] resulting in eating inferior food[12] or eating less[10]

The free-ranging dog (*Canis lupus familiaris*) is an ideal model organism to study the effect of various factors and challenges of an urban, human-dominated environment on food preference and selection, if any. Despite subsisting on a carbohydrate-rich omnivorous diet they maintain a strong preference for meat in cafeteria-type trials[13,14]. They find their preferred food using the Rule of Thumb: "If it smells like meat, eat it". This preference is also exercised while scavenging by the dogs in a more realistic scenario of finding food from garbage bins. Individual dogs were found to preferentially find and eat meat pieces from the noisy background of garbage first, as compared to other types of food using the "Sniff & Snatch" strategy[15]. They show qualities of a periscopic forager by optimizing the order of sampling and eating from resource patches (in this case, the boxes) on simultaneously encountering them[16]. This type of feeding strategy has characteristics of an optimal foraging model where the highest ranked food (in both quality and quantity) is sought out first and then less preferred food are added to the diet[17]. They show different levels of social organisation, from solitary living, to living in pairs, and living in packs[18] and thus may face similar pressures as other group-living animals. We carried out a multiple-choice task with free-ranging dog (FRD) groups to understand whether they preferentially feed on meat using the Rule of Thumb, in the presence of competitors. The objectives of this paper are to examine the feeding patterns and strategies of free-ranging dogs in groups and to compare these against those of solitary dogs in the context of the predictions of optimal foraging models. We hypothesize that presence of other group members will cause a relaxation in diet selectivity and have an effect on their foraging strategies. We hypothesized that dogs would try to minimize competition by relaxing resource selectivity, rather than entering into active competition for the most preferred food.

**Methods**

**Study Area**

The study was conducted in both urban and semi-urban habitats in and around Kolkata, India (22.5726 °N, 88.3639 °E) in areas having eateries, shops, open markets of meat, fish, and vegetables, garbage dumps and dustbins. The 22 field sites have been highlighted in the maps (Supplementary Material 1). The study was carried out during June to October 2019 and during March-April 2021. All the areas had significant human presence that encompassed a varied range of human-dog interactions. In order to avoid repeating the experiment on the same groups, we carried out the experiments in different areas on different days. This prevented any learning bias in the focal dogs.

**Experimental Protocol**

The composition of the boxes simulating dustbins and the experimental protocol followed was the same as that of Sarkar et al[15] with one exception. The experiment was carried out on groups of dogs. A group was defined as a cluster of three or more dogs with at least two of them being adults, and that were observed to be either resting together or engaging in affiliative interactions[19] with each other.

Each group was allowed to interact with the set-up for a minute, like the previous experiment, after which the boxes were removed from the vicinity of the group. The observation time window started as soon as the first dog from the group sniffed from one of the boxes. Thus, although the entire time window for the experiment was one minute, only the first responder was able to avail of it. Each box was assigned a particular identity/category (Refer to Sarkar et al. 2019, for more details). The experiment was carried out on 136 groups and compared with 68 solitary dogs (both adults and juveniles) from the previous experiment.

**Behaviours and Analysis**

Each event of sniffing and eating by an individual dog within a group and time-limit was assigned an order from 1-6. Let us consider an example where the mixed box was sniffed first, followed by protein box and then the dog ate from it and this was followed by the dog eating from the mixed box and then sniffing the carbohydrate box. In this case, the order would be assigned as SM-1; SP-2; EP-3; EM-4; SC-5, where SM stands for sniffing mixed, EP stands for eating protein and so on. The sniffing time for each dog started from the moment they began investigating by sniffing at their first box and ended at the moment the dog encountered food and picked it up in its mouth. We also noted down the following things: a) the time taken to sniff from the time the experimental set-up was available for investigation to the dogs, called latency to sniff time b) the time taken to eat from a box after its first instance of sniffing known as latency to eat time, and c) frequency of sampling before and after eating d) the total time and attempts spent on each box and the time spent in each such attempt e) whether dogs showed vigilance behaviour, defined as visual scanning of the surrounding environment[20].

We used Bayesian statistics to analyse our data. We ran multiple generalised linear mixed effects model for parameter estimation using the brms package[21] in R 4.0.2[22]. We used a weakly informative prior specifying a normal distribution (mean = 0, sd = 1) on the fixed effects and an exponential distribution (sd = 1) on the random effects for all the models (Model specific shape parameter priors have been mentioned in relevant sections). Thus, we favoured no specific direction for the effects while keeping their magnitude within reasonable values. We used 95% Bayesian Credible Interval (BCI) to assess the variability in parameter estimates and to interpret our findings[23]. We reported log odds ratios (LOR) for the logistic, multinomial, and beta regressions. Credible intervals that did not overlap zero (unless a different value is stated) indicated difference between parameters.

For models, we ran 5000 (8000, for complex models) iterations of the Markov Chain and discarded the first 1000 (or 2000) as a warm-up. We used efficient approximate leave-one-out cross-validation (loo)[24] or k-fold cross validation[25] to select the best fitted candidate model amongst the ones we fitted. There were two exceptions to the usage of this condition: a) When there was only a single candidate model b) When a model with lower loo score made more logical sense, as a better estimation of prediction error does not automatically lead to a better model[26]. Furthermore, we used graphical posterior predictive check to assess if our best fitted model gives valid predictions with respect to our observed data.

We carried out time-to-event analysis using the rstanarm package[27] to compare latencies. For each model, we compared the posterior estimate of the standardised survival curve to the Kaplan-Meier survival curve which helped us to assess the model fit to the observed data (estimated survival function check). We also validated our result using restricted mean survival time (RMST), interpreted as the average event-free (as in, no sniffing/eating) time up to a pre-specified important time point, τ. For all models Gelman and Rubin's Rhat statistic was used to assess convergence and it was adequate in all cases with Rhat = 1[28].

**Results**

**Comparison of Preference between boxes (quantified through sniffing and eating)**

We analysed the activity patterns of the dogs in groups. We quantified the total instances of sniffing and eating separately and considered whether a particular box or boxes had been sniffed/eaten more than the other. The response variable was binary ("sniff"/ "eat"), "Y" when a box was sniffed/eaten and "N" when it wasn't. We ran a logistic regression with Bernoulli distribution.

Our best fitted model was a hierarchical model with "random effect" of individual dog nested within group nested within place. "box" refers to the identity of the box sniffed/eaten by the dog. "groupid" refers to the identity of the group the dog belongs to and "id" refers to its own identity. Here activity is a placeholder for the sniff and eat activities.

activity ~ 1 + box + (1 | place/groupid/id)

The results of the regression showed that for sniffing, neither carbohydrate box (LOR: -0.29; CI: -0.66, 0.07) nor protein box (LOR: 0.17; CI: -0.21, 0.54) were sniffed any more or less than mixed box. But for eating, carbohydrate box was eaten from less as compared to mixed box (LOR: -1.48; CI: -1.99, -1.00) while eating from protein box showed no such difference (LOR: 0.03; CI: -0.36, 0.42).

**Comparison of Preference for sniffing and eating first between boxes**

We analysed whether the dogs sniffed and ate first from a particular box(es) more than the others. This response variable was trinary ("box"), "M" for mixed box, "P" for protein box, and "C" for carbohydrate box. We ran a multinomial logistic regression. Our model was a hierarchical Intercept only model with random effect of group nested within place. "groupid" refers to the identity of the group the dog belongs to.

box ~ 1 + (1 | place/groupid)

The results of the regression showed that while sniffing first, neither carbohydrate box (LOR: -0.17; CI: -0.62, 0.23) nor protein box (LOR: -0.18; CI: -0.65, 0.22) were preferred any more or less than mixed box. But while eating first, carbohydrate box (LOR: -1.47; CI: -2.49, -0.60) was less preferred than mixed box. Protein box showed no such difference as compared to mixed box (LOR: -0.14; CI: -0.60, 0.33).

**Comparison of Sniff and Snatch (SnS) strategy**

We analysed whether dogs in groups displayed SnS behaviour (sniffing a box and immediately eating from it) towards a particular box or not. The response variable was binary - "y", dummy coded as 1, if SnS was displayed and "n", dummy coded as 0, if SnS was not displayed. A dog could show one of four behaviours after sniffing from a box: (a) immediately eat from it (SnS); (b) sniff from a different box; (c) eat from a different box; (d) do nothing. Out of the four only (a) was coded as 1 while the rest were coded as 0. We ran the fitted function in brms to calculate the probabilities of each category (1 or 0) on the response scale and reported probabilities with 95% credible intervals. Credible intervals that don't overlap 0.25 indicate evidence of difference between category levels, rather than the probability of the event being the outcome of random chance. The probability value of 0.25 was used because SnS is one of four actions that the dog can take.

Our model was an Intercept only hierarchical model with random effect of group nested within place. "groupid" refers to the identity of the group the dog belongs to. The dependent variable "dfood" ("food" being a placeholder for either one of the three boxes) was a binary variable (1 or 0) indicating if a dog displayed SnS behaviour towards the respective box.

dfood ~ 1 + (1 | place/groupid)

The results showed that the probability of dogs in groups following SnS towards the mixed and protein boxes is higher than a random chance event (For mixed - mean: 0.492; 95% CI: 0.381, 0.605; For protein - mean: 0.471; 95% CI: 0.373, 0.588) and lower than a random chance event for carbohydrate box (mean: 0.133; 95% CI: 0.064, 0.230).

**Sampling of boxes**

**Sampling before Eating**

We analysed whether more than one box was sampled either by dogs in groups or individually or both before the first event of eating from a box. Sampling was initiated when the dog sniffed a different box from the one which it was currently sniffing. That was counted as Sampling Event 1, sniffing another box or returning to the previous box was Sampling Event 2, and so on. The end point was when the dog started eating. The response variable was binary ("morethanone"), "no" when only one box was sniffed at before eating or moving away from the set-up, and "yes" when more than one box was sampled. We ran a logistic regression with Bernoulli distribution.

Our best fitted model was a hierarchical model with "random effect" of group nested within place. "firstbox" refers to the first box sniffed by the dog (dummy coded as "0" for Mixed, "1" for Carbohydrate, "2" for Protein boxes) and "status" refers to whether the dog was solitary or in a group (dummy coded as "0" for group, "1" for solitary).

morethanone ~ 1 + firstbox + status + (1 | place/group)

The results of the regression showed that compared to a mixed box, sniffing from a carbohydrate box first made it more likely to sample more than one box (LOR = 0.69; 95% CI = 0.13, 1.26). Protein box showed no such difference (95% CI = -0.30, 0.82). Solitary dogs were more likely to sample more than one box, compared to dogs in groups (LOR = 0.82; 95% CI = 0.08, 1.64). The conditional effects plot of the model is given in Figure 1.

**Sampling after Eating**

We analysed whether dogs sampled from other boxes after they had initiated the first event of eating and, if yes, the number of such sampling events. We ran a multilevel hurdle model for the zeroes ("hu": whether or not sampling post eating was done) and a Poisson regression for the counts of such non-zero attempts.

Our best model was a hierarchical model with group nested within place. "morethanone" refers to whether or not the dog sampled multiple boxes before the first event of eating and "grpsize" refers to the number of group members present on the day of the experiment for a particular group (solitary dogs were a "group of 1", i.e. their group size was 1).

number ~ 1 + morethanone + grpsize + (1 | place/group),

 hu ~ 1 + morethanone + grpsize + (1 | place/group)

The results of the regression showed that dogs who did multiple pre-eating sampling were less likely to sample other boxes post-eating (mean Estimate = 0.98, 95% CI = 0.29, 1.68). Furthermore, larger the group size, more the likelihood of sampling post-eating (mean Estimate = -0.33, 95% CI = -0.59, -0.10). Either pre eating sampling or the group size do not have any effect on the rate of post-eating sampling.

**Latency in sniffing**

We defined the latency to sniff as a time-to-event variable. The time interval available for the dog to complete the task of sniffing was 300 seconds (the total duration of a trial before it is deemed unsuccessful and started when the first box was opened). We only considered the first responders in the group data, that is, the dogs who responded first to the set-up due to the fact that they resemble solitary dogs most closely in terms of accessibility and availability of resources and time. In this study we specified a proportional hazards model for the "hazard" of sniffing. For easier interpretation, the term "hazard" is replaced by "occurrence" in subsequent sections with no change in its meaning or mathematical formulae. We considered the effect of *condition,* a variable that tells us whether dogs are part of a group or solitary, on the occurrence of sniffing. The occurrence ratio (OR; $\exp(\beta_1$/coefficient for condition)) quantifies the relative increase in the occurrence that is associated with a unit-increase in the relevant covariate, whilst holding any other covariates in the model constant. In our model, OR is a time fixed quantity.

Our best fitted model was a weibull model. Here, the outcome variable is time taken to sniff after first opening a box, denoted by "secstosniff", the "censored" variable lets the model know if the dog has sniffed within the time interval of 300 seconds (1 for sniffing, 0 for not sniffing) and the predictor "condition" is a variable which can take one of two levels- solitary and group.

(secstosniff, censored) ~ condition

The results of the regression gave us the estimated ORs and showed that individuals in the solitary condition have lower rates of sniffing relative to the ones in groups. The inferred median condition effect is -0.939 with an estimate of 0.156 for the standard deviation of the marginal posterior distribution of the covariate effect. The occurrence of eating is 0.391 times lower than the occurrence of eating by individuals in group. The 95% CI of the posterior is completely below zero (-1.2, -0.6). We also checked the posterior distribution for the absolute difference in RMST between the two conditions at the average time, $\tau = 33s$ (rmst.group – rmst.solitary). The 95% CI of the difference posterior lies entirely below zero (-12.41, -6.73). This provided evidence that the RMST is higher in solitary condition, thus sniffing earlier is more likely in group.

**Latency in eating**

The time interval available for the dog to complete the task of eating was 60 seconds (the total duration of the experiment once a dog started sniffing a box). Our best fitted model was a cubic m-spline model with degrees of freedom, $df = 5$ and $\delta = 3$. The outcome variable is time taken to eat after first sniffing a box, denoted by "secstoeat", the "censored" variable lets the model know if the dog has eaten within the time interval of 60 seconds (1 for eating, 0 for not eating) and the predictor "condition" is a variable which can take one of two levels- solitary and group.

(secstoeat, censored) ~ condition

The results of the regression gave us the estimated ORs and showed that individuals in the solitary condition have lower rates of eating relative to the ones in groups. The inferred median condition effect is -0.449 with an estimate of 0.220 for the standard deviation of the marginal posterior distribution of the covariate effect. The occurrence of eating is 0.638 times lower than the occurrence of eating by individuals in group. As the 95% CI of the posterior is touching zero (-0.9, 0), we checked the posterior distribution for the absolute difference in RMST between the two conditions at the halfway time, $\tau = 30s$ (rmst.group – rmst.solitary). The 95% CI of the difference posterior lies entirely below zero (-5.11, -0.137). This provided evidence that the RMST is higher in solitary condition, thus eating earlier is more likely in group. The posterior probability that RMST at $\tau = 30$ is higher for solitary than group is 0.981 giving further support to our result. We plotted the predicted survival function between 0 and 60s for a dog in each of the condition in Figure 2.

**Vigilance Behaviour**

The following behaviours were included under vigilance behaviour : a) alert scanning while eating or suspending foraging b) continuously looking at something while eating or suspending foraging c) showing wariness while looking at something (including experimenters/humans) d) following the movement of traffic, humans or other moving objects even if dog has moved away from boxes e) alert and scanning while sitting

The following behaviours were not considered: a) looking at insects/walls/ground/boxes b) looking at humans in a positive gesture (for e.g. affiliative) c) If eyes are not clearly visible and head movement is negligible d) sitting and looking with a relaxed posture

Our best model was a hierarchical logistical model with group nested within place. "scanenvdn" is a binary variable and refers to whether a dog showed vigilant behaviour and "status" refers to whether the dog is solitary or in group.

scanenvdn ~ 1 + status + (1 | place/groupid)

The results of the regression showed that compared to dogs in groups, solitary dogs are more likely to scan their environment (LOR = 1.26; 95% CI = 0.53, 2.02). The conditional effects plot of the model is given in Figure 3.

**Total handling attempts**

Total handling attempts (THA) is defined as the number of times a particular box has been engaged (sniffed/ foraged/ eaten from) with over the entirety of time available to the dog of a minute. An attempt is initiated when a dog interacts with a box through one of the activities mentioned above and ends when the dog stops interacting with the box. Disengagement happened when a dog started sniffing another box, started looking around for potential dangers or opportunities or was distracted due to external factors like car horns or ticks. We analysed whether dogs handled a box and, if yes, the number of such handling attempts. We ran a multilevel hurdle model for the zeroes ("hu": whether or not handling was done) and a Poisson regression for the counts (number: number of handling attempts for the non-zero responses).

Our best model was a hierarchical model with individual dog ("id") nested within group, nested within place. "tha" refers to total handling attempts and "box" refers to the identity of the box

tha ~ 1 + box + (1 | place/group), hu ~ 1 + box + (1 | place/group)

The results of the regression showed that the identity of the box did have an effect on the rate of THA, with THA rate for the carbohydrate box lower as compared to mixed box (mean Estimate = -0.27, 95% CI = -0.50, -0.03). The rate of THA was not different between protein and mixed box. Furthermore, the identity of the box had no effect on the likelihood of being handled.

**Total handling time (THT)**

THT for a particular box is defined as the proportion of time spent by a dog throughout the totality of its handling events on a particular box out of its total available time to forage. We ran a zero-one-inflated beta distribution to account for the zeroes and ones in the dataset. We reported log odds ratios with 95% credible intervals. We computed the log odds difference between the three box types to check if one box type was handled more or less than the other.

Our candidate model was a hierarchical model with individual dog ("id") nested within group, nested within place. "propht" refers to the proportion of the time a dog spent at a box foraging (handling a resource patch) out of the total available time to forage and "box" refers to the identity of the box the dog spent that time in and "status" refers to whether they were in group or solitary.

propht ~ 0 + box: status + (1 | place/group/id)

phi/zoi/coi ~ 0 + box: type + (1 | place/group/id),

The results of the regression showed that dogs in groups spent lesser proportion of time handling the carbohydrate box as compared to mixed and protein boxes as shown by the posterior difference (95% CI for M – C = 0.051, 0.240; 95% CI for P – C = 0.086, 0.301). Solitary dogs spent greater proportion of time handling protein box as compared to carbohydrate box (95% CI for P – C = 0.010, 0.321) but no such difference between protein and mixed boxes (95% CI for P – M = -0.285, 0.014) or between mixed and carbohydrate boxes (95% CI for M – C = -0.125, 0.178) was observed. No difference across type (group vs solitary) was seen for the same box (e.g., group mixed vs solitary mixed). The conditional effects plot of the model is given in Figure 4.

**Unit handling time**

Unit handling time is defined as the amount of time spent in each handling attempt out of the available proportion of handling time. It is formulated as handling time divided by THA. We carried out a generalized linear regression with zero-one-inflated beta distribution to account for the zeroes and ones in the dataset.

Our best model was a hierarchical model with individual dog ("id") nested within group, nested within place. "prophand" refers to the proportion of the time a dog spent at a box handling it out of the available proportion of handling time, "box" refers to the identity of the box the dog spent that time in and "type" refers to whether they were in group or solitary.

prophand ~ 1 + box + type + (1 | place/group/id),

$$\text{phi/zoi/coi} \sim 1 + \text{box} + \text{type} + (1 \mid \text{place/group/id})$$

The results of the regression showed that solitary dogs spent lesser proportion of time in each handling attempt as compared to dogs in groups (LOR= -0.35, 95% CI = -0.61, -0.09). Furthermore, dogs spent lesser proportion of time in each handling attempt for carbohydrate box as compared to mixed box (LOR = -0.74, 95% CI = -0.97, -0.51). Handling time of protein and mixed boxes showed no such difference.

**Discussion**

Resources are often distributed in a heterogenous and stochastic fashion across time and space in patches. The acquiring of such resources is dependent on the ability to make profitable foraging decisions by foragers. Such decisions can only be made by gathering information about patch heterogeneity. Patch quality information can be gathered while exploiting it. This is known as sampling for information[29]. Foragers harvest information from the environment by sampling and investigating through visual, olfactory, and chemical cues and using such information to modify their behaviour accordingly[30,31]. In uncertain environments, animals spend a considerable amount of their time investigating, with a significant proportion of that spent in resource rich patches[30,32]. As the resources in a patch deplete, foragers may have to frequently reassess patch choice and make decisions whether to stay, move on or re-visit a previous patch[33,34].

In our experiment, the boxes simulating dustbins acted as heterogeneous resource patches about which the dogs had no prior knowledge. Thus, the only way to collect information and assess the quality of the patches was through sampling them. A major decision node around which we split the dogs' sampling efforts was the first eating event. Previous research had shown that solitary dogs harvested the best resource patch first but showed no such preference during sniffing. Although this hinted at an optimal foraging strategy, it was not immediately clear whether dogs carefully assessed available patches before making a decision or they behaved impulsively or were able to hone on the best available resource. Our current experiment showed that sampling played an important role during foraging in FRDs. Solitary dogs sampled multiple "patches" (boxes), leaving behind sub-optimal food as evidenced by the fact that sampling likelihood increased if dogs came across carbohydrate box first, till they encountered the best available resource and then ate from it. This demonstrates that dogs were able to assess and evaluate the quality and quantity of resources available in a patch and take this into account before making a decision of eating. On the other hand, dogs in groups were more likely to eat from the first box they encountered but would frequently sample other "patches" while harvesting resource from their current "patch". This post-sampling tendency increased with an increase in group size, thus hinting at an implicit effect of competition, in the presence of conspecifics, at play. Additionally, in both cases, solitary, and group, no particular box was sniffed first more than the others thus demonstrating that the dogs did not have an instantaneous clue of the location of preferred food and thorough sampling was the only way for them to gather information about the "patches". Overall, dogs who showed pre-eat-sampling behaviour were less likely to do post-sampling elucidating that the behaviour was not random but rather had the specific purpose of finding the best available resource patch and the application of the behaviour varied according to the level of competition.

In keeping with the unpredictable environment that they live in and the fluctuating food resources available therein, FRDs were found to investigate and engage with all the available resource "patches" (boxes) though they concentrated their maximal efforts on the energy-rich boxes (protein and mixed) choosing to handle them more number of times than the energy-poor boxes (carbohydrate). Furthermore, instead of allocating *all their time* to the best "patch" once located, the dogs allocated different proportions of their available time to different "patches" depending on their quality with better "patches" allocated more time but suboptimal patches allocated a small proportion of time too. Although this seems to be a deviation from optimal behaviour, it is hypothesized that the dogs are actually prioritizing a long-term adaptation to a fluctuating environment, akin to titmice foraging[34]. In their natural habitat, food resource clumps such as dustbins and garbage dumps vary in food abundance across time such that high abundance food sources can become low abundance ones at a later time and vice versa. In such a scenario, it is advantageous to spend some time handling all "patches", especially if the distance between the "patches" are minimal, as is the case here, in order to continually track the status of the environment and change the allocation as and when necessary. Additionally, dogs spent a lower proportion of time in each handling attempt for carbohydrate box as compared to mixed and protein box. This can be because of two reasons. One, apart from handling carbohydrate box fewer times, the smaller time allocation for each attempt shows that dogs expend less effort for a suboptimal resource. The other reason seems to be that between feeding from protein and mixed boxes, dogs would sample from carbohydrate box too and would quickly assess the value to be lower than that of the other two boxes and move back to them. Unfortunately, we were unable to observe the amount of food eaten by individual dogs. Thus, we were unable to find out the point of resource depletion at which the dogs shifted to another patch.

FRDs are known to forage singly as well as in groups[18]. Group living animals have to balance between access to quality nutrition and conspecific competition. Decreased resource selectivity in the presence of conspecifics reflects an adaptive response to competition. Such decreased selectivity can also be observed in the FRDs in the current experiment. While solitary dogs sought to maximise both quality and quantity of the available food by preferentially feeding from the protein box[15], dogs in groups show a relaxation in their preference by feeding from both protein and mixed boxes equally. While it might be hypothesized that the dogs moved to mixed box only after they believed that the protein box had no valuable resource to offer them, we don't think that is the case here. The dogs ate first from both the boxes equally showing that instead of protein box being the top "patch" in preference hierarchy, the dogs have relaxed their choice to include both as equally preferable. As dogs ate chicken pieces in overwhelmingly high amount than bread (80.5% of the total food eaten), it implies that even while eating from the mixed box, they selectively ate the chicken pieces over bread. Thus, while there is increased acceptance in terms of feeding from lower valued "patches" (lower quantity of preferred food), there is little change in acceptance of lower valued food as compared to solitary dogs. Solitary dogs follow the SnS strategy exclusively for the protein box whereas the dogs in groups apply the strategy to both mixed and protein boxes. Taken together, these results show that even in groups, dogs do follow the Rule of Thumb. These results lend credence to optimal foraging prediction that states that if foods of higher value are available, low value items should be rejected, regardless of their abundance[17]. Similarly, FRDs seem to sequester the preferred food, meat, first followed by bread. This seems to be an adaptation of the scavenging lifestyle that these dogs lead wherein they try to maximise nutrients in any form they can while maintaining their preference.

FRDs are known to be bolder in groups[35]. The reduced time demonstrated by groups in responding to the experimental set-up and in eating as compared to solitary dogs imply the presence of such traits during foraging. Indeed, it was found that solitary dogs are more likely to indulge in vigilance behaviour as compared to those in groups, perhaps due to the likelihood of facing competition from other dogs and scavengers while foraging. Since this was not the focus of this experiment, we did not attempt to look into this behaviour in great detail. Dogs in groups are also known to spend more time engaging with the boxes in each foraging attempt. Foraging in groups confer multiple benefits to animals, some of which such as reduced vigilance[36] and more time spent in foraging[37] seem to be behind the dogs' action here. However, we must specifically investigate other factors that might elicit similar behaviour in future studies. Intragroup competition and social rank appropriate behaviour might be two such factors[38].

In this study, we investigated the foraging behaviour of FRDs in groups and compared it against that of solitary dogs. We found that the foraging tactics of dogs fulfil multiple characteristics of an optimal foraging strategy and that these animals are highly plastic in their behaviour and adaptable in their strategies, as befits an organism living in an unpredictable environment. Further research may focus on finding out the energy threshold that make a "patch" valuable and how animals assess that. Studies may seek to explore the mechanism behind the inverse relationship between group size and vigilance and disentangling the various factors that come into play when foraging in group such as competition, hierarchy and food density. Examining the effect of these confounding factors in tandem and alone will shed greater insight into the foraging behaviour of these animals and the causal mechanism behind such behaviours as reduced latency.

**Figures**

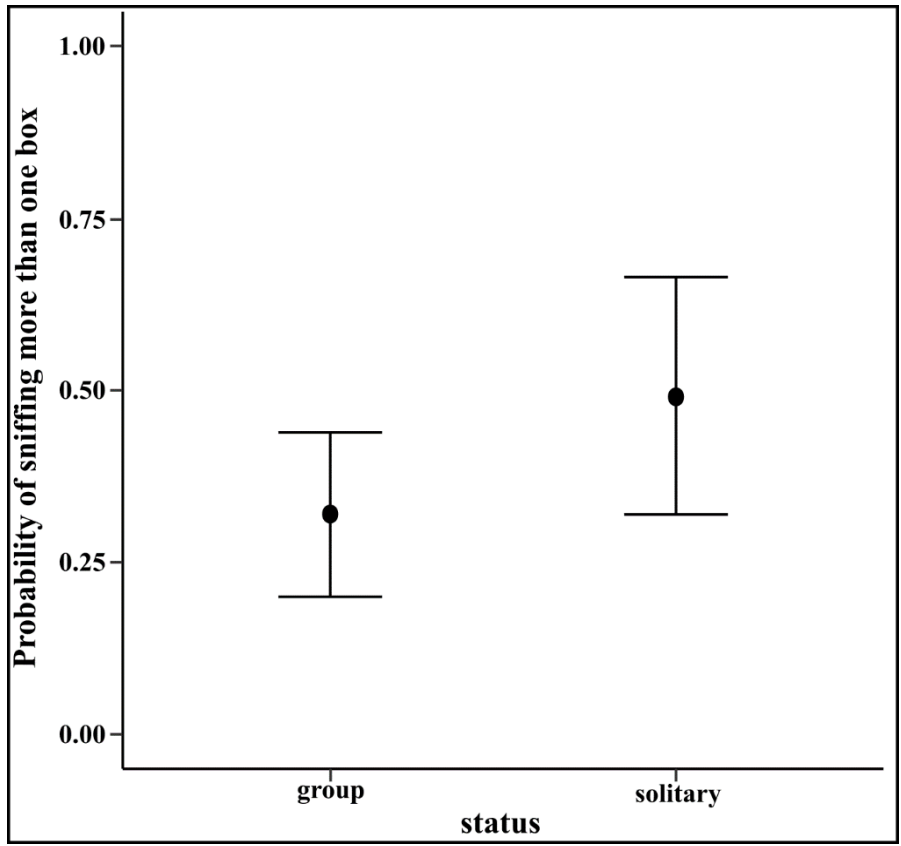

Fig 1: Conditional effects plot of the pre-eating sampling logistical model; the error bars display the 95% credible intervals; the circles represent posterior medians

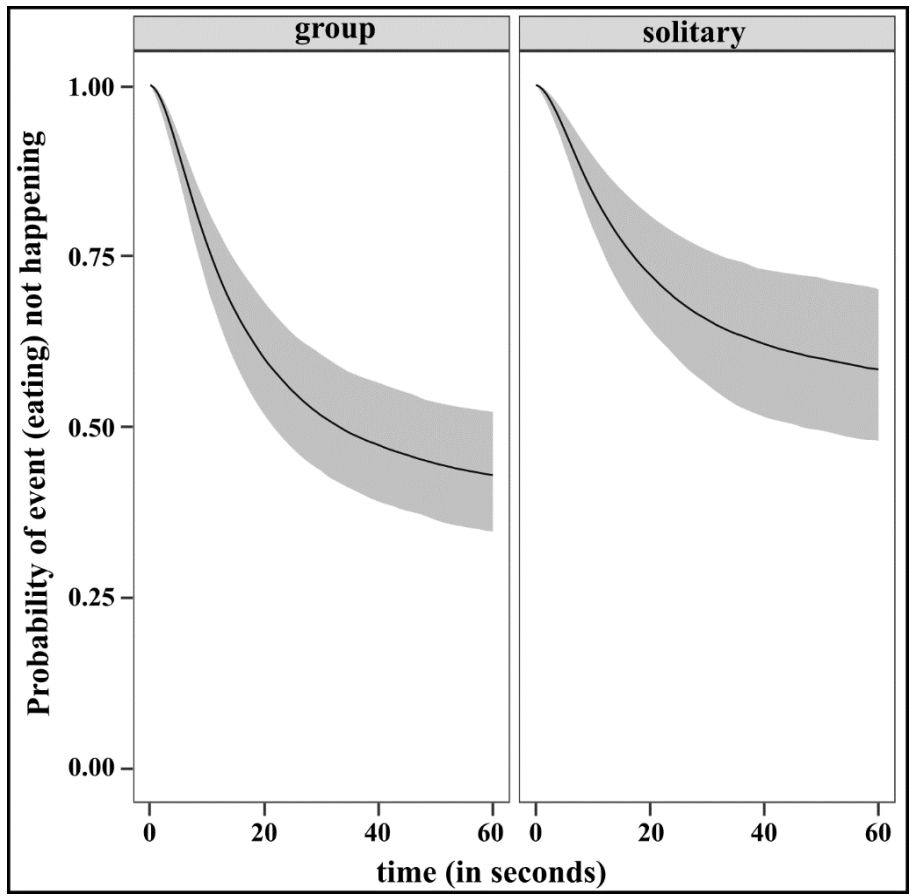

Fig 2: Predicted survival (no eating) function (posterior median and 95% uncertainty limits) for an individual dog in either Solitary or Group condition

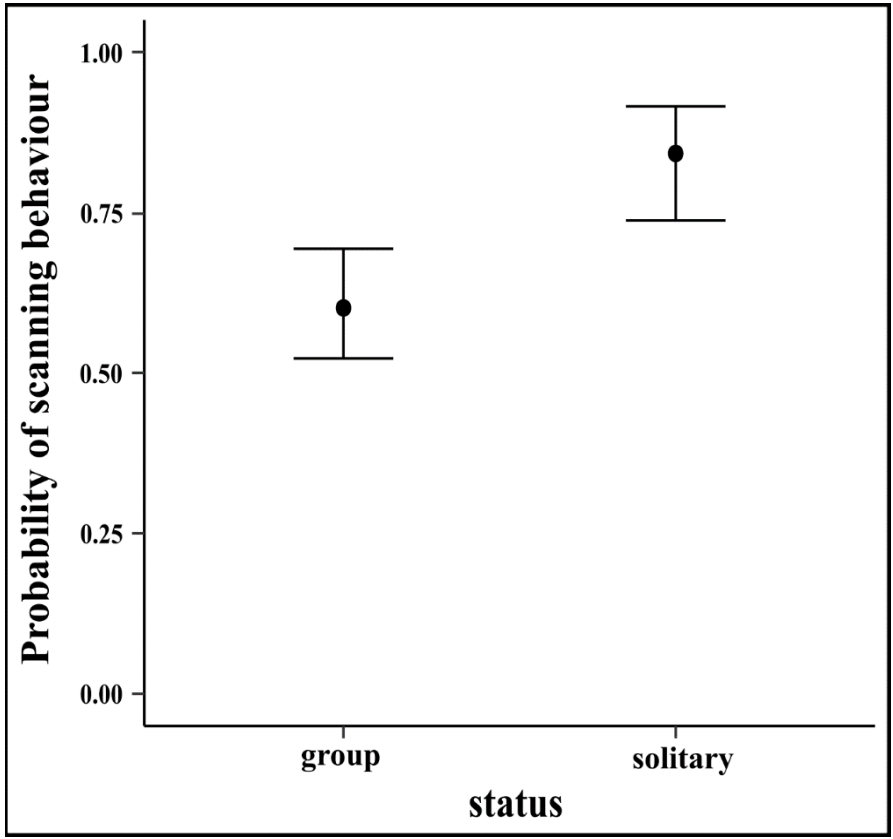

Fig 3: Conditional effects plot of the vigilance logistic regression; the error bars display the 95% credible intervals; the circles represent posterior medians

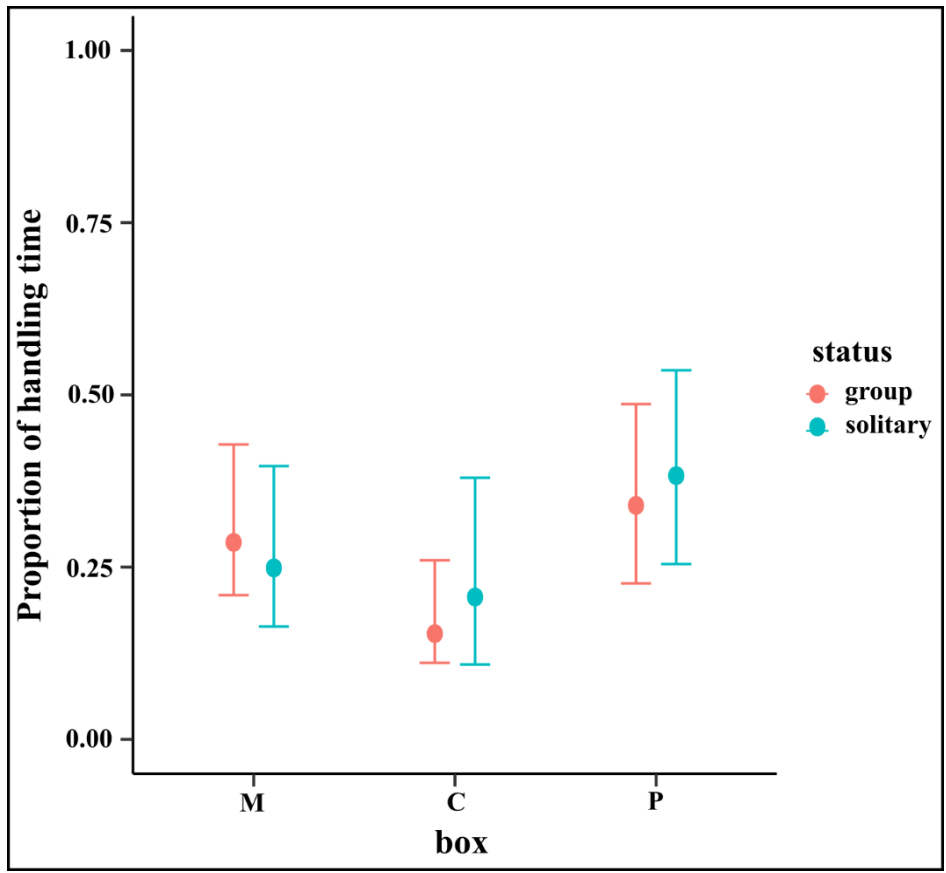

Fig 4: Conditional effects plot of the zero-one-inflated beta regression model; the error bars display the 95% credible intervals; the circles represent posterior medians

**Acknowledgement**

RS would like to thank Dr. Scott Claessens (School of Psychology, University of Auckland, New Zealand) for his immense help, guidance and mentoring of RS in Bayesian statistics, model building and for providing manuscript inputs. Without his patience and kindness this paper would not have been possible. RS would also like to thank Dr. Satyaki Mazumder (Department of Mathematics and Statistics, IISER-Kolkata, India) and Narayan Srinivasan (Department of Mathematics and Statistics, IISER-Kolkata, India) for their valuable inputs. RS would like to express gratitude to Animal Behaviour Collective for their support during the covid pandemic.

**Funding**

RS was supported by IISER Kolkata Institute fellowship. This work was supported by the Science and Research Board, Department of Science and Technology, Government of India.

**Supplementary Materials**

**Supplementary material 1**

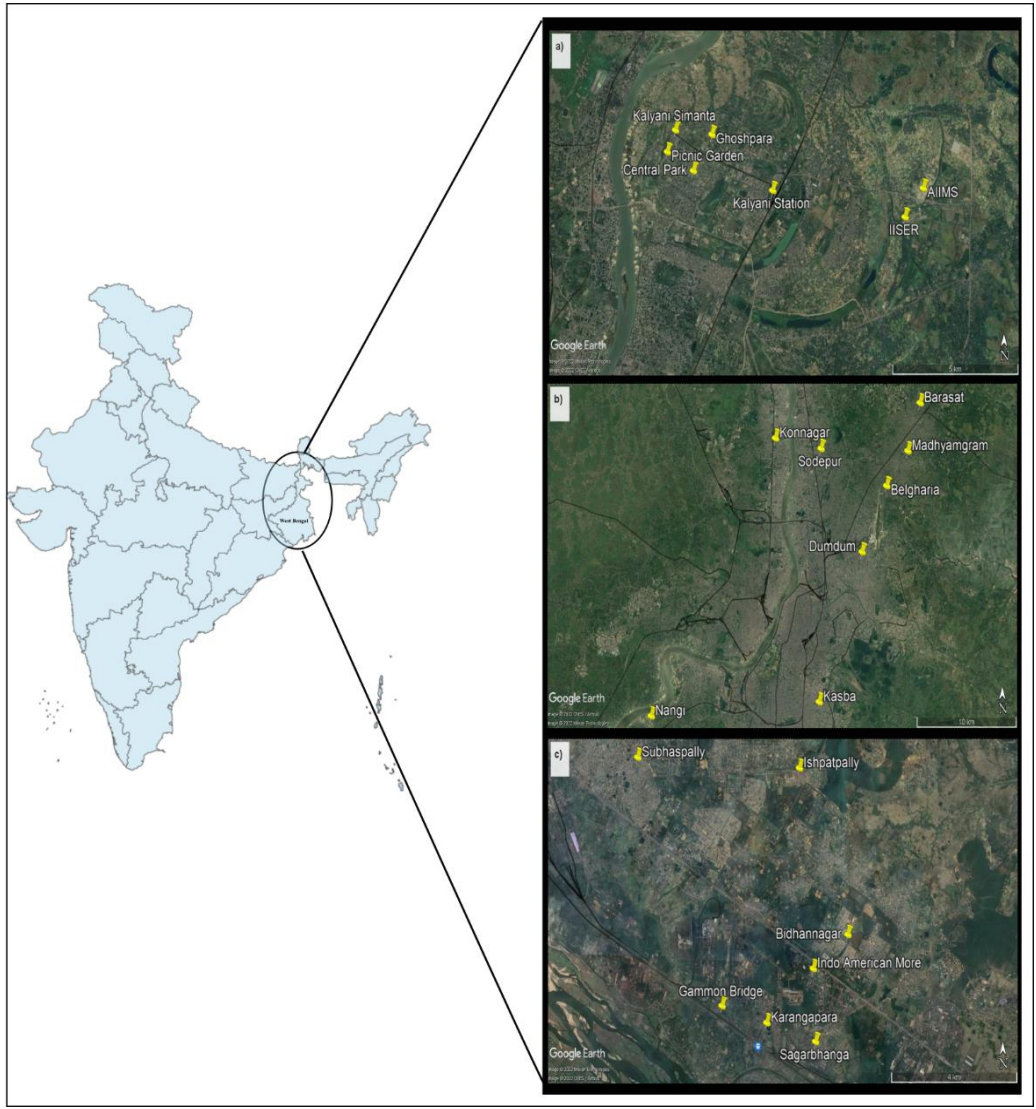

*S_fig 1: Map of India which highlights the area of West Bengal in which the study was conducted: (a) Kalyani (b) Kolkata and surrounding suburbs (c) Durgapur*

**Supplementary material 2**

**Food eaten**

We carried out a logistic regression with zero-inflated binomial distribution to account for the excess zeroes in the dataset. We used kfold cross- validation (CV) to select the best fitted model amongst the ones we fitted. We reported log odds ratios with 95% credible intervals. We computed the log odds difference between the two food types to check if one food type was eaten more than the other.

Our best fitted model was a hierarchical model with "varying" effects of group nested within place. We used a prior beta(2, 2) for "zi" to regularize it towards 0.5. The outcome variable was "Count" which referred to the number of food pieces eaten per group out of a total of 15 pieces, denoted here by trials (n). "foodType" refers to the type of food available to the dog for eating (dummy coded as "0" for bread, "1" for chicken). "participants" denoted the number of dogs eating in a group.

Count | trials (15) ~ 1 + foodType + participants + (1 + foodType | place/group)

The results of the regression showed that compared to bread, chicken increased the likelihood of proportion of food items being eaten (LOR = 2.54, 95% CI = 1.19, 3.65). The number of participants had a positive effect on the likelihood of food items being eaten (LOR = 1.92, 95%CI = 1.61, 2.27). The density plot of the model parameter estimates have been shown in Supplementary figure 2 below. The density of the group intercept "random effect" (sd_place: group_Intercept) and the random slope of foodType (sd_place : group_foodType) on it are both away from zero validating our choice of including them in the model. Clearly, group has an effect on the likelihood of food eaten and the effect of food type varied between groups. This, as the "participants" parameter shows, is probably because the number of participants in each group is different and groups with larger number of participants eat more food or when only a single member of a group eats during the experiment, it eats only one type of food. Another possibility is that different members eat different food, dependent on availability. The posterior distribution of the place Intercept contained zero, elucidating that it had no effect.

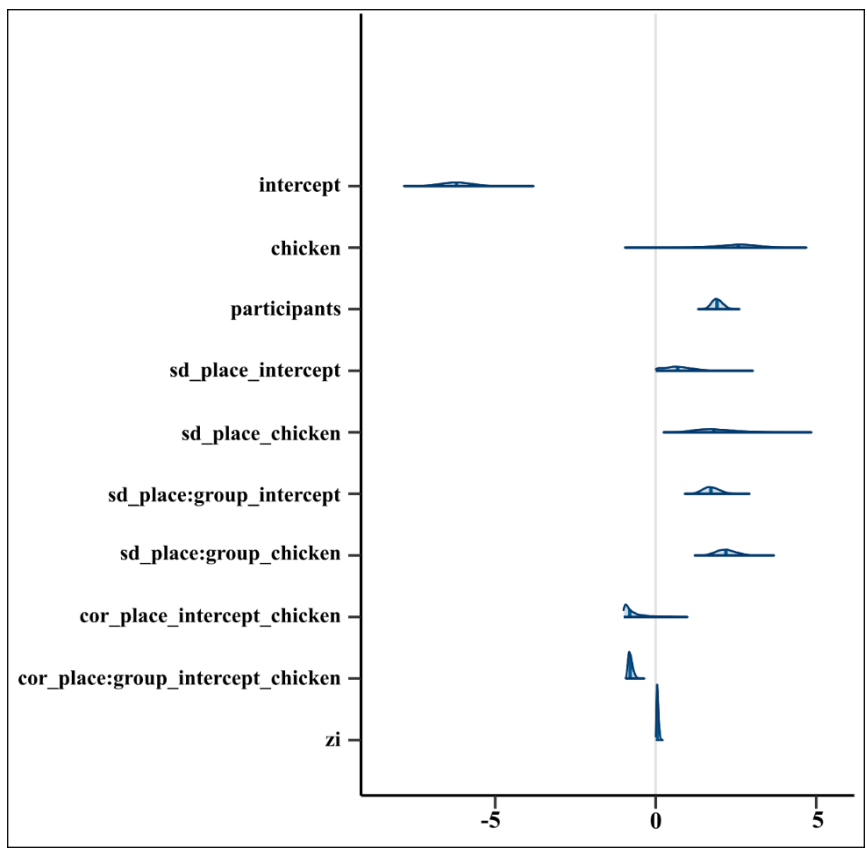

*S_fig 2: Density plot of the model parameter estimates*